\documentclass[aps,showpacs,jcp,superscriptaddress,floatfix,twocolumn,citeautoscript]{revtex4-1}
\usepackage{graphicx}
\usepackage{amsmath}
\usepackage{amssymb}
\usepackage{bm}
\usepackage{dcolumn}
\usepackage{subfigure}
\usepackage{units}
\usepackage{hyperref}
\usepackage[usenames]{color}
\usepackage{booktabs}
\usepackage{multirow}
\usepackage{siunitx}
\usepackage{wrapfig}
\usepackage{lipsum}
\usepackage{setspace}
\hypersetup{pdfborder=0 0 0,colorlinks=true,citecolor=blue,linkcolor=blue}
\input epsf

\newcommand{\BN}{\textit{h}-BN}

\begin{document}

\title{Coulomb drag in anisotropic systems: a theoretical study on a double-layer phosphorene}
\author{S. Saberi-Pouya}
\author{T. Vazifehshenas}
\email{t-vazifeh@sbu.ac.ir}
\affiliation{Department of Physics, Shahid Beheshti University, G. C., Evin, Tehran 1983969411, Iran}
\author{M. Farmanbar}
\affiliation{Faculty of Science and Technology and MESA$^{+}$ Institute for Nanotechnology, University of Twente, P.O. Box 217, 7500 AE Enschede, The Netherlands}
\author{T. Salavati-fard}
\affiliation{Department of Physics and Astronomy, University of Delaware, Newark, DE 19716, USA}

\begin{abstract}
We theoretically study the Coulomb drag resistivity in a double-layer electron system with highly anisotropic parabolic band structure using Boltzmann transport theory. As an example, we consider a double-layer phosphorene on which we apply our formalism. This approach, in principle, can be tuned for other double-layered systems with paraboloidal band structures.  Our calculations show the rotation of one layer with respect to another layer can be considered a way of controlling the drag resistivity in such  systems. As a result of rotation, the off-diagonal elements of drag resistivity tensor have non-zero values at any temperature. In addition,  we show that the anisotropic drag resistivity is very sensitive to the direction of momentum transfer between two layers due to highly anisotropic inter-layer electron-electron interaction and also the plasmon modes. In particular, the drag anisotropy ratio, $\rho^{yy}/\rho^{xx}$, can reach up to $\thicksim 3 $ by changing the temperature. Furthermore, our calculations suggest that including the local field correction in dielectric function changes the results significantly. Finally, We examine the dependence of drag resistivity and its anisotropy ratio on various parameters like inter-layer separation, electron density, short-range interaction and insulating substrate/spacer.

\end{abstract}
\date{\today}
\maketitle

\maketitle
\section{Introduction}

The advent of two dimensional (2D) materials has sparkled a considerable scientific interest due to their unique properties and their potential for applications in electronic devices. Atomically thin 2D materials, such as graphene \cite{Geim:nat13}, monolayer black phosphorous (phosphorene)\cite{Ling:pnas15}, hexagonal boron-nitride\cite{Xu:chemrev13}, and the transition-metal dichalcogenides (TMDs)\cite{Liu:ChemS15} represent a particularly interesting class of 2D materials including both semiconductors and metals. Phosphorene which is an interesting monatomic layered crystalline material, can be mechanically exfoliated from the bulk black phosphorus due to the weak van der Waals interaction between layers\cite{Likai:nat14}.Unlike in group IV elemental
materials such as graphene, silicene\cite{Grazianetti:2d16} , or germanene \cite{Acun:jpcm15,Cahangirov:prl09}, phosphorene is a semiconductor with puckered orthorhombic structure. This semiconductor has a nearly direct bandgap and its band structure shows a large anisotropy and high sensitivity to deformation \cite{Rodin:prl14}. This fact suggests that crystallographic properties play an important role in the electronic behavior of this system. Recent studies reveal a high degree of anisotropic electronic and optical properties for the phosphorene, which further confirms the importance of this new 2D semiconductor as a promising candidate for electronic\cite{liu:acs14,Sa:nano15}, thermoelectric\cite{liu:nano15}, and plasmonic applications. \cite{Low:prl14,Xia:nat13,Engel:nanol14,Rudenko:prb14,Low:prb14}
\begin{figure}
	\includegraphics[width=8.5cm]{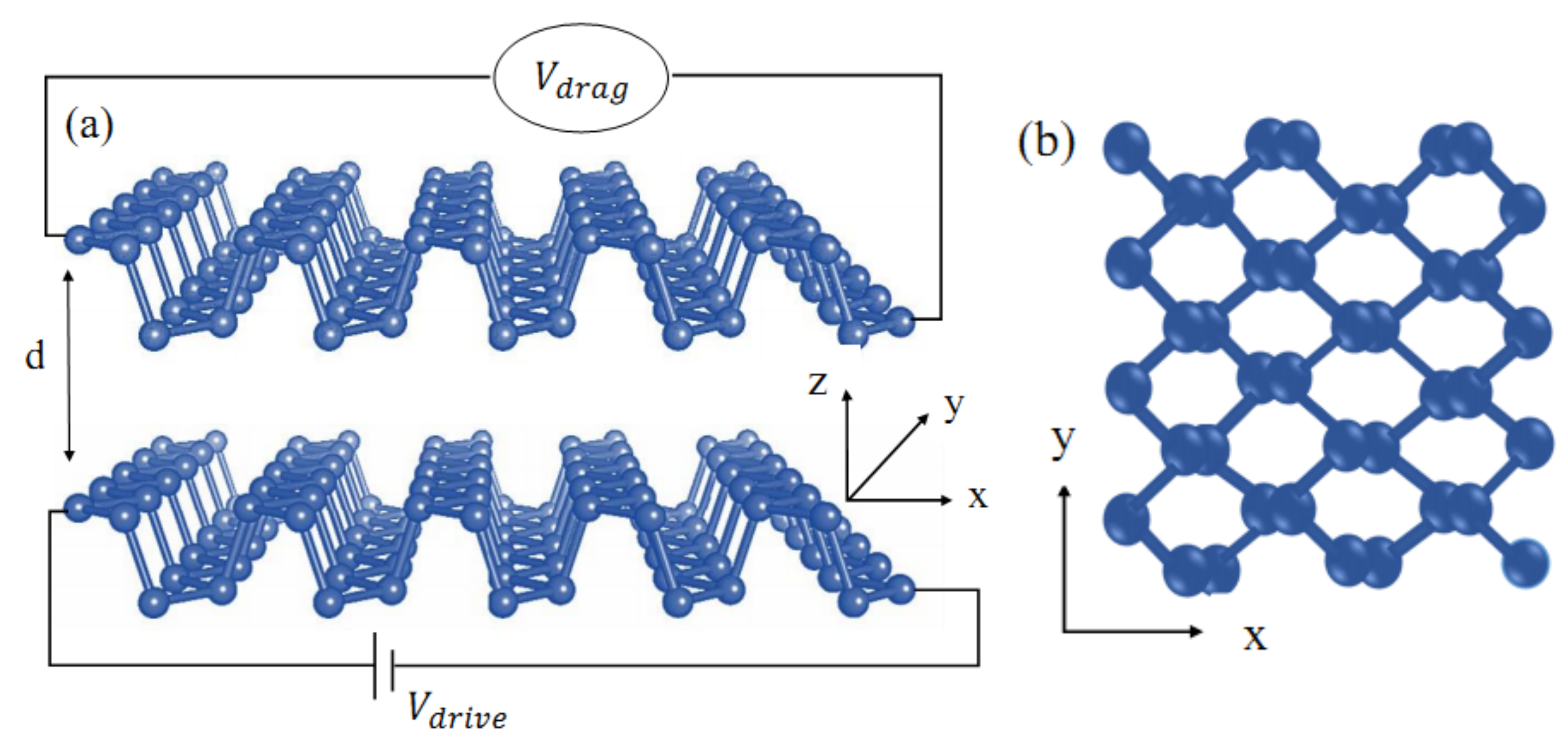}
	\caption{(Color online)(a) Side view of a double-layer phosphorene system with the separation of $d$ in the drag setup. (b) Top view of phosphorene.}
	\label{fig1:phos}
\end{figure}

Double-layered 2D structures consisting of two parallel electron or hole systems which are kept in a close vicinity demand special attention due to many-body and transport properties \cite{Shulenburger:nanol15,Vazifehshenas:pla10,Berdiyorov:jap15,Roldan:prb13}. The inter-layer Coulomb interaction plays a significant role in these correlated systems. In Coulomb drag phenomenon, momentum can be transferred from interacting electrons in one layer to electrons in the adjacent layer\cite{Narozhny:arx15,Yurtsever:ssc03,Tanatar:prb00,Vazifehshenas:pe07,Vazifehshenas:pb15,Gamucci:nat14}. The momentum transfer takes place through inter-layer Coulomb interaction, but does not involve any carrier exchanges. This phenomenon has been previously studied in a few nanostructures such as double quantum wells (DQW)\cite{Flensberg:prl94,Narozhny:arx15,Jauho:prb2002,Solomon:prl89,Noh:prb98,Tso:prl92,Jauho:prb93,Lian:prb93,Weber:nat05,Gramila:prl91,Croxall:prl08}.

 Van der Waals bonding of 2D heterostructures makes it possible to spatially separate two layers of graphene (or any other 2D materials) down to  several nanometers, by inserting a few atomic layers of a 2D insulator, for instance \BN, to isolate the layers from one another. As shown by Gorbachev\cite{Gorbachev:natphys12}, $et.al.$, a double-layer graphene system has a strong Coulomb drag resistivity with respect to GaAlAs heterostructures\cite{Gorbachev:natphys12,Tse:prb07,Hwang:prb11,Carrega:njp12}. Likewise, Phosphorene is a 2D material, however, with highly anisotropic energy dispersion in contrast to graphene and other 2D materials. Therefore, it is particularly worthwhile to examine the effect of anisotropy on the Coulomb drag \cite{Ezawa:jpcs15}. 

In this paper we investigate the effect of band anisotropy on the Coulomb drag resistivity in a double-layer electron system, consisting of two individual isolated layers which are coupled via Coulomb interaction. We start from the expression for drag resistivity based upon the semiclassical Boltzmann transport equation theory in the relaxation time approximation, and develop a general formalism, which includes the effect of anisotropic energy dispersion and rotationally misaligned 2D systems. As an example we apply this formalism to calculate Coulomb drag resistivity in a double-layer phosphorene system(see figure \ref{fig1:phos}). Numerical results show a strong drag resistivity dependence on the band anisotropy. It indicates that the drag resistivity along bigger mass, i.e., $m_{y}$, has a larger value. Furthermore, we discuss how the drag resistivity and its anisotropy ratio depend on the carrier density, inter-layer separation, rotation of layers and the choice of insulating substrate/spacer. The drag resistivity is enhanced substantially in the $y$ direction when  the Hubbard local field correction (LFC) is added to our formalism; LFC includes the short-range exchange effect between electrons in the same the layer. 

The rest of paper is organized as follows. Section \ref{sec:Model}, describes the model, theoretical formalism, and the Coulomb drag resistivity in 2D anisotropic systems. In Sec. \ref{sec:B}, the dynamically screened inter-layer potential and also temperature-dependent anisotropic polarization function are presented. Section \ref{sec:C} contains results for a double-layer phosphorene and the summary and conclusions are presented in Sec. \ref{sec:final}.

\maketitle
\section{Coulomb drag resistivity in 2D anisotropic systems}
\subsection{Model}
\label{sec:Model}
We consider a system composed of two parallel spatially separated 2D electron gases with anisotropic parabolic-like band structures. In this system the carriers are coupled through Coulomb interaction and there is no tunneling between layers so the Fermi energies and chemical potentials can be considered independently. The inter-layer Coulomb interaction can cause momentum transfer from the electrons in the drive layer, layer 1, to the carriers in the drag layer, layer 2. In doing so it generates a potential difference across the layers. In Fig.\ref{fig1:phos}(a) we show a schematic of the drag setup with phosphorene layers as anisotropic 2D electron gases. Also, a top view of phosphorene monolayer can be seen in Fig.\ref{fig1:phos}(b). 

 The drag (inter-layer) resistivity, $\rho$, can be defined as \cite{Flensberg:prb95a}:
 \begin{equation}
\sum\limits_{\alpha=x,y}\rho_{21}^{\alpha \beta}J_{1,\alpha}=\Xi_{2,\beta}, 
\label{eq:mod1}
\end{equation}
 where $\mathbf{J}$ and $\mathbf{\Xi}$ are the current density and electric field, respectively. $\alpha$ and $\beta$ indexes label $x$ and $y$ components and $\rho_{21}^{\alpha \beta}=\rho_{12}^{\beta \alpha}$. The drag resistivity in the linear regime has been studied in a variety of theoretical approaches assuming parabolic and non-parabolic band structures (particularly graphene with a linear energy spectrum) and considering both momentum-dependent and independent intra-layer relaxation times\cite{Jauho:prb2002,Solomon:prl89,Narozhny:prb12,Kats:prb11,Bruno:JPCM12,Carrega:njp12}. Using different theoretical approaches such as the memory function formalism\cite{Lian:prb93}, Kubo formula based on the leading-order diagrammatic perturbation theory \cite{Flensberg:prb95} and the linear response Boltzmann transport equation, the drag resistivity matrix has been obtained with the assumption of momentum-independent intra-layer transport time (see Appendix): 
\begin{equation}
\begin{aligned}
\rho_{21}^{\alpha \beta}={}&\frac{\mathrm{\ensuremath{\hbar}}}{2\pi e^{2}n_{1}n_{2}k_{B}T}\int\frac{d^{2}q}{(2\pi)^{2}}q_{\alpha}q_{\beta}\\   
			    &\times\int_{0}^{\infty}d\omega\frac{|U_{21}(\mathbf{q},\omega)|^{2}\Im\Pi_2(\mathbf{q},\omega)\Im{\Pi_1(\mathbf{q},\omega)}}{\sinh^{2}(\hbar \omega/2k_{B}T)}
\label{eq:mod2}
\end{aligned}
\end{equation}
 Here, $U_{21}(\mathbf{q},\omega)$ is the temperature-dependent dynamically screened inter-layer interaction, $\Pi_{i}(\mathbf{q},\omega)$ and $n_{i}$ being the 2D non-interacting polarization function and electronic density of $i$th layer, and k$_{B}$ is the Boltzman constant. In Sec. \ref{sec:C}, we will rewrite Eq.(\ref{eq:mod2}) to make it more convenient to use in a double-layer electron gas system with anisotropic band structure.

\subsection{Inter-layer potential and temperature-dependent anisotropic polarization function }
\label{sec:B}
The dynamically screened inter-layer potential can be obtained by solving the corresponding Dyson equation \cite{Badalyan:prb12}:
\begin{equation}
U_{ij}(\mathbf{q},\omega)=\frac{V_{ij}(q)}{det|\epsilon_{ij}(\mathbf{q},\omega)|},
\label{eq3}
\end{equation}
where $V_{ij}(q)=\nu(q)\exp(-qd(1-\delta_{ij}))$ is the unscreened 2D Coulomb interaction with $d$ being the layer spacing. $\nu(q)=2\pi e^{2}/q\kappa$, with $\kappa$ being the average dielectric constant. Finally $\epsilon_{ij}(\mathbf{\mathbf{q}},\omega)$ is the dynamic dielectric matrix of the system. For systems with high electron density it is reasonable to employ the RPA to calculate $\epsilon_{ij}(\mathbf{\mathbf{q}},\omega)$\cite{Hwang:prb09,Wen:nanotech12}:
\begin{equation}
\epsilon_{ij}(\mathbf{q},\omega)=\delta_{ij}+V_{ij}(q)\Pi_{i}(\mathbf{q},\omega)
\label{eq:epsil}
\end{equation}
At low electron densities, the short-range local field effects are not negligible and must be included in the dielectric matrix by replacing  $(1-G_{ij}(q))V_{ij}(q)$ for, $V_{ij}(q)$ where $G_{ij}(q)$ denotes the static intra- ($i=j$ ) and  inter-layer ($i \neq j$ ) elements of LFC matrix, respectively. Here we incorporate only the intra-layer components of the LFC factor because of their stronger effect on the drag resistivity\cite{Jonson:jpc76}:
\begin{equation}
G_{ii}(q)=\frac{q}{2\sqrt{q^2+k_{F}^2}},
\label{eq5}
\end{equation}
where $G(q)$ and $k_{F}=\sqrt{2\pi n}$ is  the Hubbard LFC factor and the Fermi wave vector, respectively, with $n$ being the electron density. For an electron gas system, the non-interacting polarization function can be obtained from the following equation\cite{Bohm:pr53}:
\begin{equation}
\Pi_{i}(\mathbf{q},\omega)=-\frac{g_{s}}{\nu} \sum \limits_{\mathbf{k}} \frac{f^{0}(E_{\mathbf{q}}^{i})-f^{0}(E_{\mathbf{k+q}}^{i})}{E_{\mathbf{q}}^{i}-E_{\mathbf{k+q}}^{i}+\hbar\omega+i\eta} 
\label{eq6}
\end{equation}
Here $f^{0}(E_{\mathbf{q}}^{i})$ is the Fermi distribution function in layer $i$ at energy $E$ corresponding to the wave vector $\mathbf{q}$, $g_{s}=2$ is spin degeneracy and $\eta$ is the broadening parameter, which accounts for disorder in the system. The temperature-dependent dynamic polarization function for intra-band transition in an anisotropic 2D material can be calculated by making use of the following anisotropic parabolic energy dispersion relation
\begin{equation}
E_{\mathbf{k}}^{i}=\frac{\hbar^2}{2}(\frac{k_{x}^{2}}{m_{x}}+\frac{k_{y}^{2}}{m_{y}})-\mu_{i},
\label{eq:energy}
\end{equation}
in Eq.(\ref{eq6}) for the polarization function:
\begin{equation}
\begin{aligned}
\frac{\Pi_{i}(\mathbf{q},\omega)}{g_{2d}}={}&-\int dK \frac{\Phi_{i}(K,T)}{Q}\Bigg[sgn(\Re(Z_{-}))\frac{1}{\sqrt{Z_{-}^{2}-K^{2}}}\\
							    &-sgn(\Re(Z_{+}))\frac{1}{\sqrt{Z_{+}^{2}-K^{2}}}\Bigg]
\end{aligned}
\end{equation}
In the above symmetric form of temperature-dependent anisotropic polarization function, we define $\mathbf{Q}=\sqrt{m_{d}/\hat{M}}(\mathbf{q}/k_{F})$, $\mathbf{K}=\sqrt{m_{d}/\hat{M}}(\mathbf{k}/k_{F})$ where $\hat{M}$ is the mass tensor with diagonal  element $m_{x}$ and $m_{y}$ along $x$ and $y$ direction, and  $m_{d}=\sqrt{m_{x}m_{y}}$ is the 2D density of state mass. Moreover, $g_{2d}=m_{d}/\pi \hbar^{2}$ and $Z_{\pm}=((\hbar\omega+i\eta)/\hbar Q \nu_{F})\pm(Q/2)$ with $\nu_{F}=\hbar k_{F}/m_{d}$ and $\Phi_{i}(K,T)$ is given by:
\begin{equation}
\Phi_{i}(K,T)=\frac{K}{1+\exp[(K^{2}E_{F}^{i}-\mu_{i})/k_{B}T]}
\end{equation}
where $\mu_{i}$ is the chemical potential of layer $i$, which is determined by the particle number conservation condition \cite{Ashcroft}:
\begin{equation}
\mu_{i}+k_{B}T\ln[1+\exp(-\mu_{i}/k_{B}T)]=E_{F}^{i}
\end{equation}
Here, we consider $\mathbf{q}=q(\cos\theta,\sin\theta)$ , in accordance to the notation in Ref. \onlinecite{Rodin:prb15}, to introduce rotational parameter for the layers. Rotational angle, $\tau_{i}$, is defined as the angle between $x$-axis in the laboratory frame and $x$ direction of the $i$th layer. So, we can write $Q=q\sqrt{m_{d}R_{i}}/k_{F}$   in which the orientation factor, $R_{i}$ , is expressed as:
\begin{gather}
R_{i}=\bigg(\frac{\cos^{2}(\theta-\tau_{i})}{m_{x}}+\frac{\sin^{2}(\theta-\tau_{i})}{m_{y}}\bigg)
\end{gather} 
In case of double-layer phosphorene, we have $m_{x}\approx 0.15m_{0}$ and $m_{y}\approx 0.7m_{0}$  where $m_{0}$ is the free electron mass \cite{Low:prb14}. As it is well known, electronic collective modes of a double-layer system are obtained from zeros of the dielectric function determinant, Eq.(\ref{eq:epsil}). In the presence of intra-band single particle excitations, there are two plasmon modes: the so-called acoustic and optical modes, which show linear $\omega_{ac}(q)\sim\sqrt{(R_{1}R_{2}/(R_{1}+R_{2})d}q$  and square-root $\omega_{op}\sim\sqrt{(R_{1}+R_{2})q}$ behavior at small wave vectors, respectively, and dependence on the orientation factors \cite{Rodin:prb15}. 
\begin{figure}
\includegraphics[width=9.0cm]{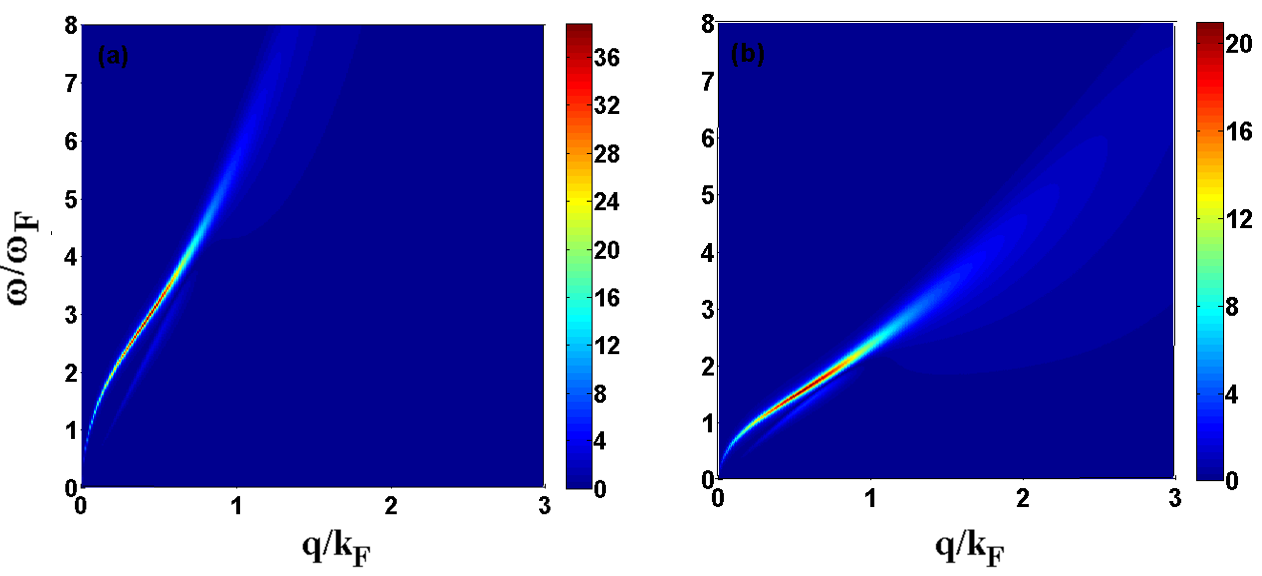}
\caption{(Color online) Loss function, $|\Im(1/det \epsilon(\mathbf{q},\omega,T)|$ , for two crystallographic directions (a) $\theta=0$ and (b) $\theta=\pi/2$ at T=100 K with $d=$5 nm , $n= 3\times10^{12}$cm$^{-2}$ and $\eta=$1 meV.}
\label{fig:2}
\end{figure}

To make the above discussion clearer, we show the loss function of a system comprising two parallel monolayer phosphorene separated by $d=$ 5 nm  at T= 100 K for two main crystallographic directions: $\theta=0$  and $\theta=\pi/2$, in Fig.\ref{fig:2}, respectively. One may notice that the acoustic plasmon mode calculated here is weaker than the optical one. This can be explained by the fact that the coherence of acoustic mode is quickly vanished due to the thermal and disorder broadening effects because the $\omega$-$q$ spectrum of this mode is very close to the single particle excitation region. Additionally, it can be recognized that the lower-energy acoustic plasmon and higher-energy optical plasmon modes following different asymptotic behavior at small wave vectors in both panels of Fig.\ref{fig:2}. Due to the anisotropic band structure, the long-lived plasmon modes disperse differently in such a way that the larger effective mass along $y$ leads to smaller resonance frequencies\cite{Rodin:prb15,Low:prl14,Jin:prb15}.

As we mentioned earlier, the RPA is reliable for systems with high electron densities. The density parameter $r_{s}=\sqrt{2}/(k_{F}a_{B}^{*})$   with effective Bohr radius $a_{B}^{*}=\kappa /(e^{2}m_{d})$, which is defined as the average distance between electrons in a non-interacting 2D electron gas, gives a measure for reliability of the RPA. In this figure, we consider the same electron density in layers, $n_{1}=n_{2}=3\times 10^{12}$cm$^{-2}$, the substrate and spacer to be Al$_{2}$O$_{3}$ with $\kappa \approx 12$ that leads to an $r_{s}\approx 1.7$  for which RPA predicts electronic screening qualitatively well. However, Hubbard LFC can improve the result of calculations by including exchange hole around interacting electrons. Later on, we will employ LFC on top of RPA to calculate the drag resistivity.

It is worth pointing out that considering a very thin double-layer phosphorene system when, at the same time, we assume there is no tunneling between the two layers, is not actually a problematic consideration because the space between layers is filled by a slim dielectric material. Al$_{2}$O$_{3}$ and \BN\ have been successfully used as a substrate and spacer to make such thin heterostructures with no inter-layer tunneling\cite{kim:scirep15,Xiaolong:natc15,Hui:acs15,Luo:ieee14}. Throughout this paper, we assume the substrate is a thick layer of the same material as spacer. 
\subsection{Drag effect in a double-layer phosphorene}
\label{sec:C}
In this section, we first derive a formula for the drag resistivity of a 2D anisotropic double-layer system with parabolic band structure and then solve it by making use of numerical methods. Eq.(\ref{eq:mod2}) is the general formula for drag resistivity based on the linearized Boltzmann transport equation. In the case of two coupled anisotropic layers, the off-diagonal components of drag resistivity tensor may have non-zero values as a result of finite $\tau$, unlike the isotropic systems such as double-layer graphene and conventional 2D electron gas. To make the difference more explicit, we rewrite Eq.(\ref{eq:mod2})  as follows:
\begin{equation}
\rho_{D}^{\alpha \beta}=\frac{\hbar^2}{(2\pi)^{3} e^{2}n_{1}n_{2}k_{B}T}\int{dq}\int_{0}^{\infty}d\omega F^{\alpha \beta}(q,\omega,T)
\label{eq12}
\end{equation}

	\begin{figure}
		\includegraphics[width=9.0cm]{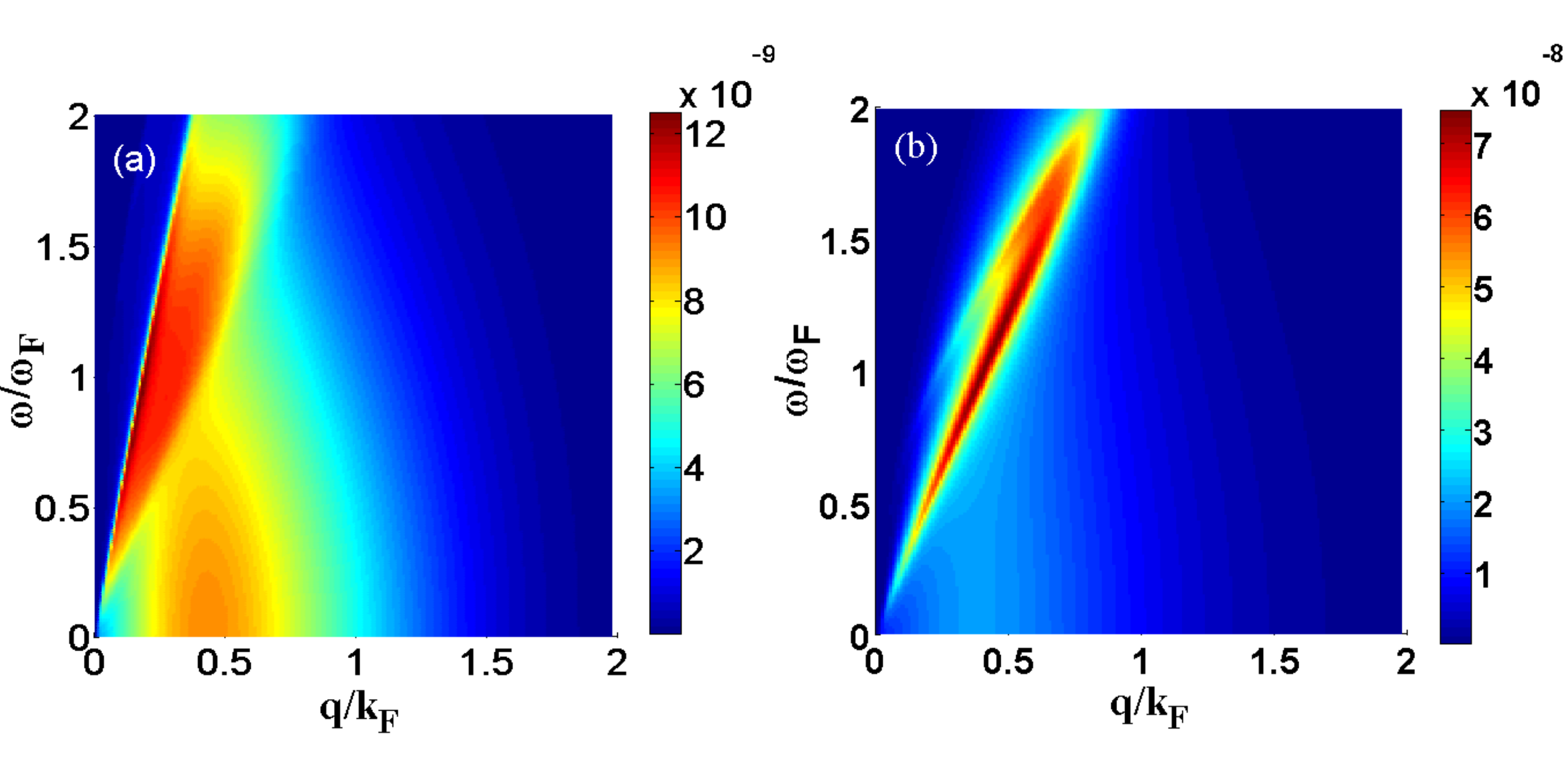}
		\caption{(Color online) (a) $F^{xx}(q,\omega,T)$ and (b) $F^{yy}(q,\omega,T)$ for two aligned  phosphorene monolayers sandwiched by Al$_{2}$O$_{3}$ layers at T=100 K with $d$= 5nm, $n = 3\times10^{12}$cm$^{-2}$ and $\eta=$1 meV .}
		\label{fig3}
	\end{figure}
where $\rho_{D}^{\alpha \beta}=\rho_{21}^{\alpha \beta}$ and $F^{\alpha \beta}(q,\omega,T)$ is defined as:
	\begin{figure}
		\includegraphics[width=8.5cm]{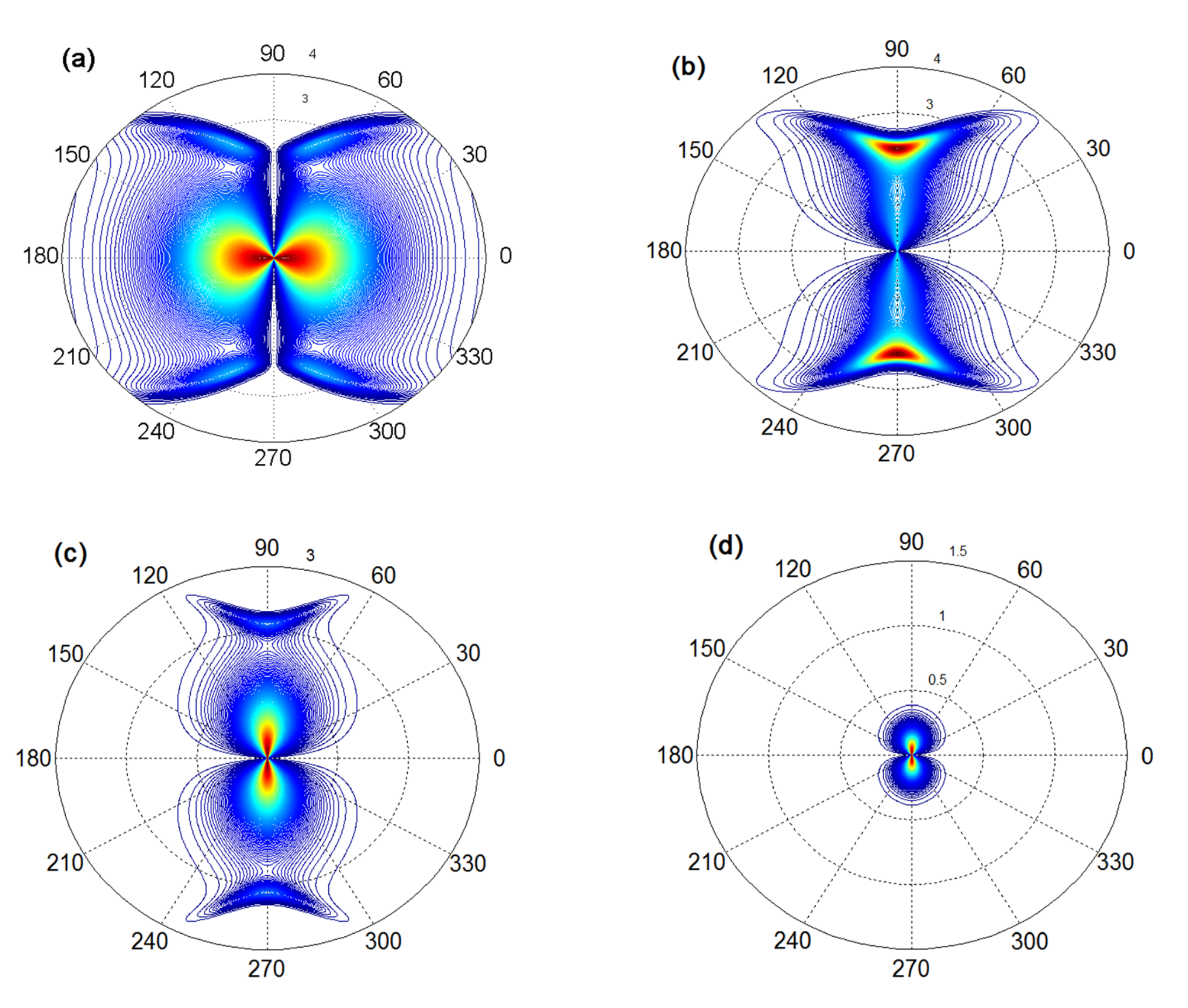}
		\caption{(Color online) The integrand of Eq.(\ref{eq13}) at $q=k_{F}$ for two aligned monolayers sandwiched by Al$_{2}$O$_{3}$ layers with $n = 3\times10^{12}$cm$^{-2}$, and $\eta=$1 meV, $d=$5 nm, along (a) $x$  and (b) $y$ directions at T=100 K and along $y$ direction at (c) T= 50 K  and (d) T= 10 K. The radial and azimuthal coordinates are $\omega/\omega_{F}$ and the angular orientation of $q$, respectively.}
		\label{fig4}
	\end{figure}
\begin{widetext}
\begin{equation}
F^{\alpha \beta}(q,\omega,T)=\int_{0}^{2\pi}d\theta \psi^{\alpha \beta}(\theta,\tau_{1},\tau_{2}) \frac{q^{3}}{\sinh^{2}( \hbar \omega/2 k_{B}T)} |U_{21}(q,\omega,T;\theta,\tau_{1},\tau_{2})|^{2}\Im\Pi_{1}(q,\omega,T;\theta,\tau_{1})\Im{\Pi_{2}(q,\omega,T;\theta,\tau_{2})}
\label{eq13}
\end{equation}
\end{widetext}
with $\psi^{\alpha \beta}$ given by
\begin{equation}
    \psi^{\alpha \beta}(\theta,\tau_{1},\tau_{2})=
    \begin{cases}
      \cos(\theta-\tau_{1})\cos(\theta-\tau_{2}), & \alpha=\beta=x \\
      \sin(\theta-\tau_{1})\sin(\theta-\tau_{2}), & \alpha=\beta=y \\
      \cos(\theta-\tau_{1})\sin(\theta-\tau_{2}), & \alpha=x,\beta=y \\
    \end{cases}
    \label{eq14}
\end{equation}

In order to understand how anisotropy affects the drag resistivity, it is worth looking into the integrand of Eq.(\ref{eq12}), $F^{\alpha \beta}(q,\omega,T)$ in more depth. In Fig.\ref{fig3}, we show $F^{\alpha \beta}(q,\omega,T)$ for a coupled system composed of two aligned phosphorene monolayers separated by 5nm at T=100 K. We use the dimensionless variables $q/k_{F}$ and $\omega/\omega_{F}$, where $\omega_{F}=\hbar^{-1}E_{F}$. Note that the integrand has significant weight in the $0<q<k_{F}$ interval, as is the case in conventional 2D electron gas \cite{Flensberg:prb95} but its values are larger along the $y$ direction. This is due to the greater effective mass of electrons in $y$ direction, which results in lower energies of the collective modes, and in this manner enhances the plasmons’ contributions \cite{Flensberg:prb95,Flensberg:prl94}. 

The angular orientation of $\mathbf{q}$ impacts the drag resistivity behavior considerably. We depict the integrand of Eq.(\ref{eq13}) along  $x(\alpha=\beta=x )$ Fig.\ref{fig4}(a) and $y(\alpha=\beta=y)$ Fig.\ref{fig4}(b) directions in an aligned-layers system. At intermediate temperature T$\sim$100 K, both modes (acoustic and optical) take part and the single particle excitation spectrum is sufficiently broadened to contribute effectively. As can be observed in the figure, the larger magnitude of the integrand occurs around $\theta=0$  and $180^{\circ}$ along the $x$ direction and around $\theta=90^{\circ}$ and $270^{\circ}$ for $y$ direction, respectively.
The integrand of Eq.(\ref{eq13}) along the $y$ direction is plotted in Fig.\ref{fig4}(c) and (d) for two different temperatures: (c) T=50 K and (d) T=10 K. According to this figure, at T=10 K the drag resistivity is mainly influenced by the acoustic mode which is lower in energy ($\omega<0.5 \omega_{F}$ ) and the optical mode contribution starts to appears at 50 K. Here, the radial and azimuthal coordinates denote $\omega/\omega_{F}$ and the angular orientation of $\mathbf{q}$, respectively.
\begin{figure}
\includegraphics[width=8.3cm]{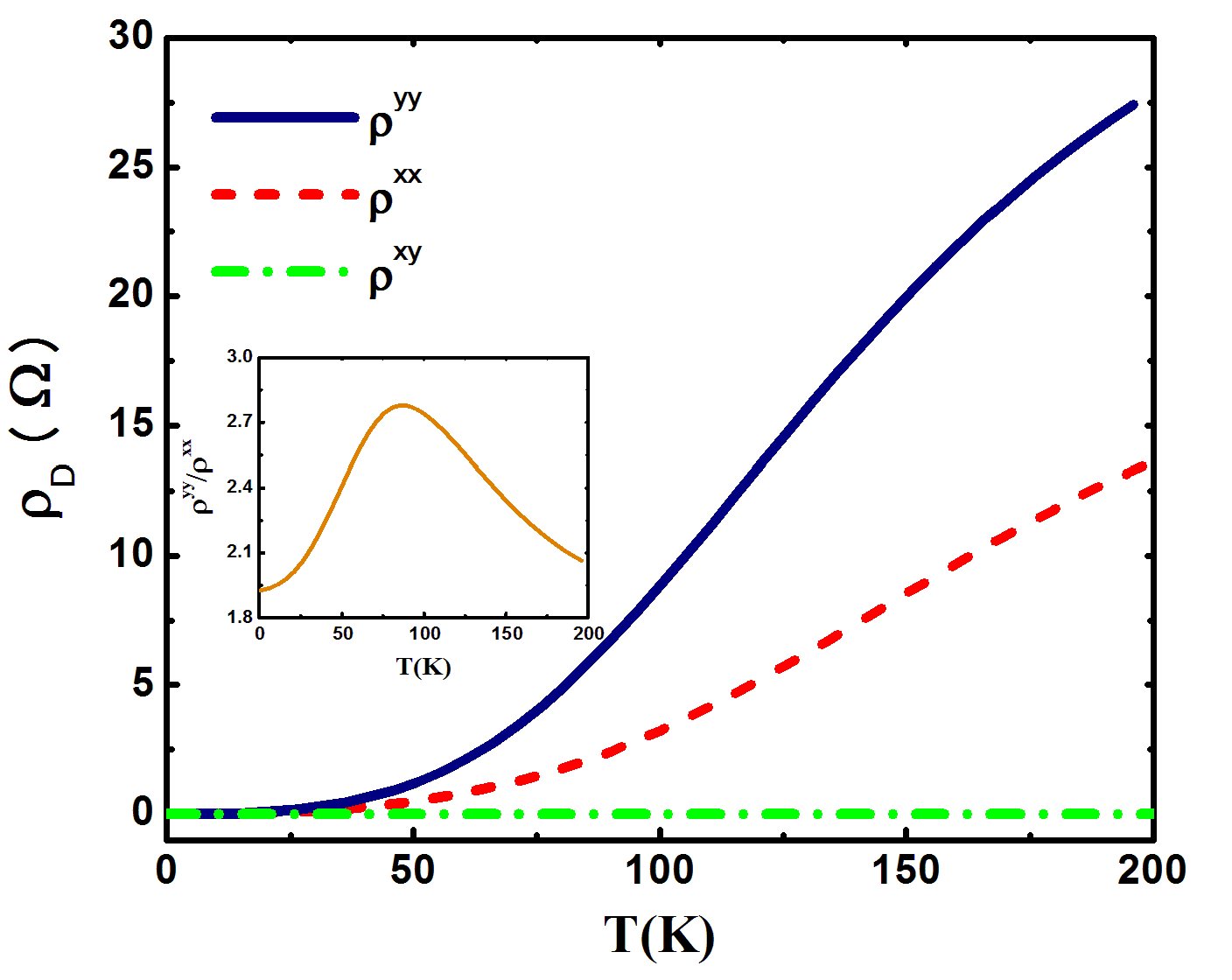}
\caption{(Color online) Anisotropic drag resistivity components, $\rho_{D}^{\alpha \beta}$ calculated within RPA as a function of temperature for two aligned phosphorene monolayers sandwiched by Al$_{2}$O$_{3}$ layers with $n = 3\times10^{12}$cm$^{-2}$ at $\eta=$1 meV and $d=$5 nm. The inset graph shows the anisotropy ratio $\rho^{yy}/\rho^{xx}$.}
\label{fig5}
\end{figure}
The first set of calculations of the drag resistivity in a double-layer phosphorene is presented in Fig.\ref{fig5}. Here we show the diagonal and off-diagonal elements of the drag resistivity tensor calculated within the RPA, versus temperature for two parallel aligned phosphorene monolayers sandwiched by Al$_{2}$O$_{3}$ layers and separated by a distance of $d=$5 nm. While the diagonal drag resistivity matrix elements increase in similar manner with temperature, there are significant differences between the values. Drag resistivity along the $x$ direction, $\rho^{xx}$, is smaller than the drag resistivity along the y direction, $\rho^{yy}$, at any temperatures of interest with a drag anisotropy ratio (see inset graph), $\rho^{yy}/\rho^{xx}$, which approximately changes from 2 up to about 3. We believe that a higher-energy resonance along $x$ direction resulting from the smaller effective mass, as discussed before, accounts for this behavior. Moreover, as it is expected from general symmetry arguments, the off-diagonal elements have zero values for aligned layers. 
\begin{figure}
\includegraphics[width=8.3cm]{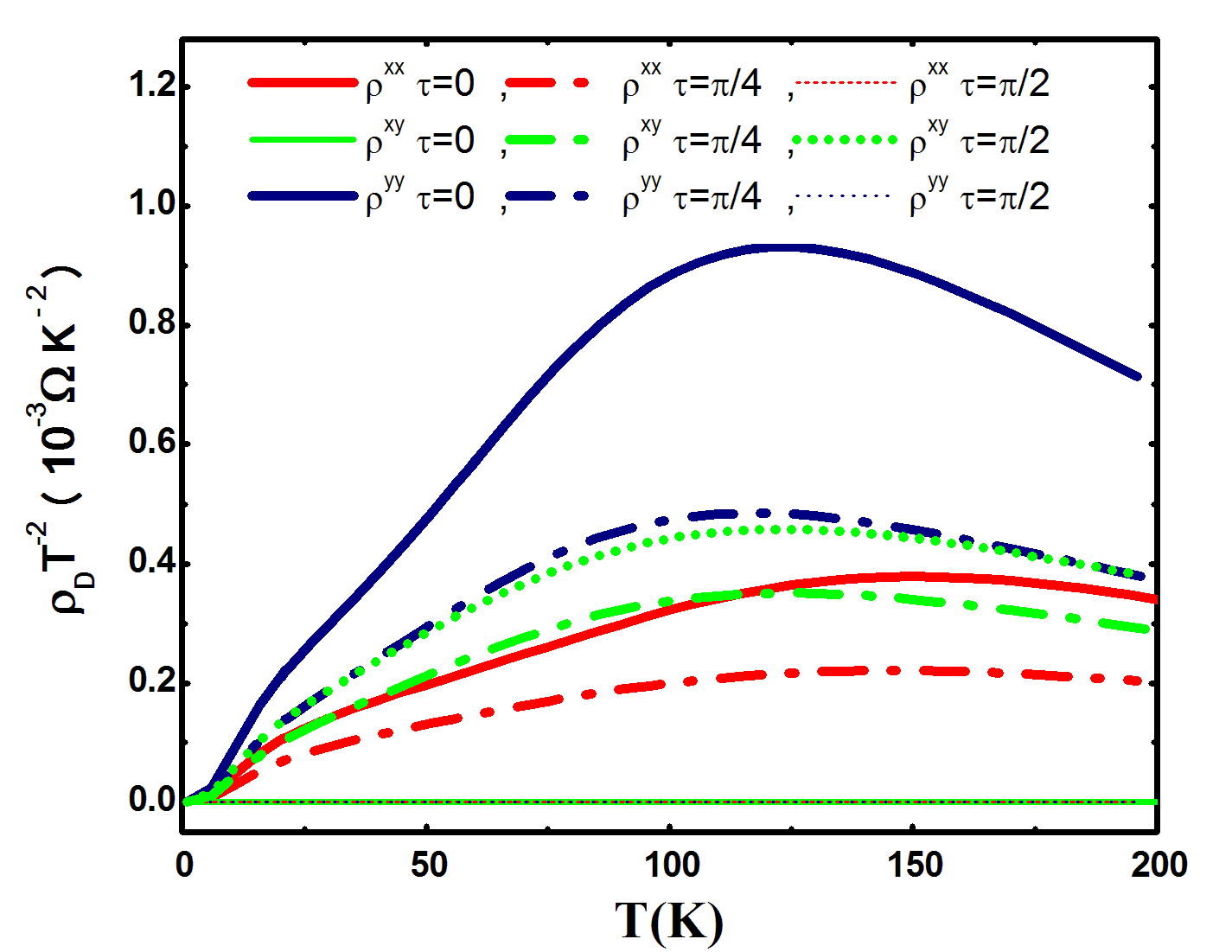}
\caption{(Color online) Scaled drag resistivity $\rho_{D}T^{-2}$ as a function of temperature for different angels with $n = 3\times10^{12}$cm$^{-2}$, and  $\eta=$1 meV, $d=$5 nm. The phosphorene layers are sandwiched by Al$_{2}$O$_{3}$ layers.}
\label{fig6}
\end{figure}
\begin{figure*}
	\includegraphics[width=12cm]{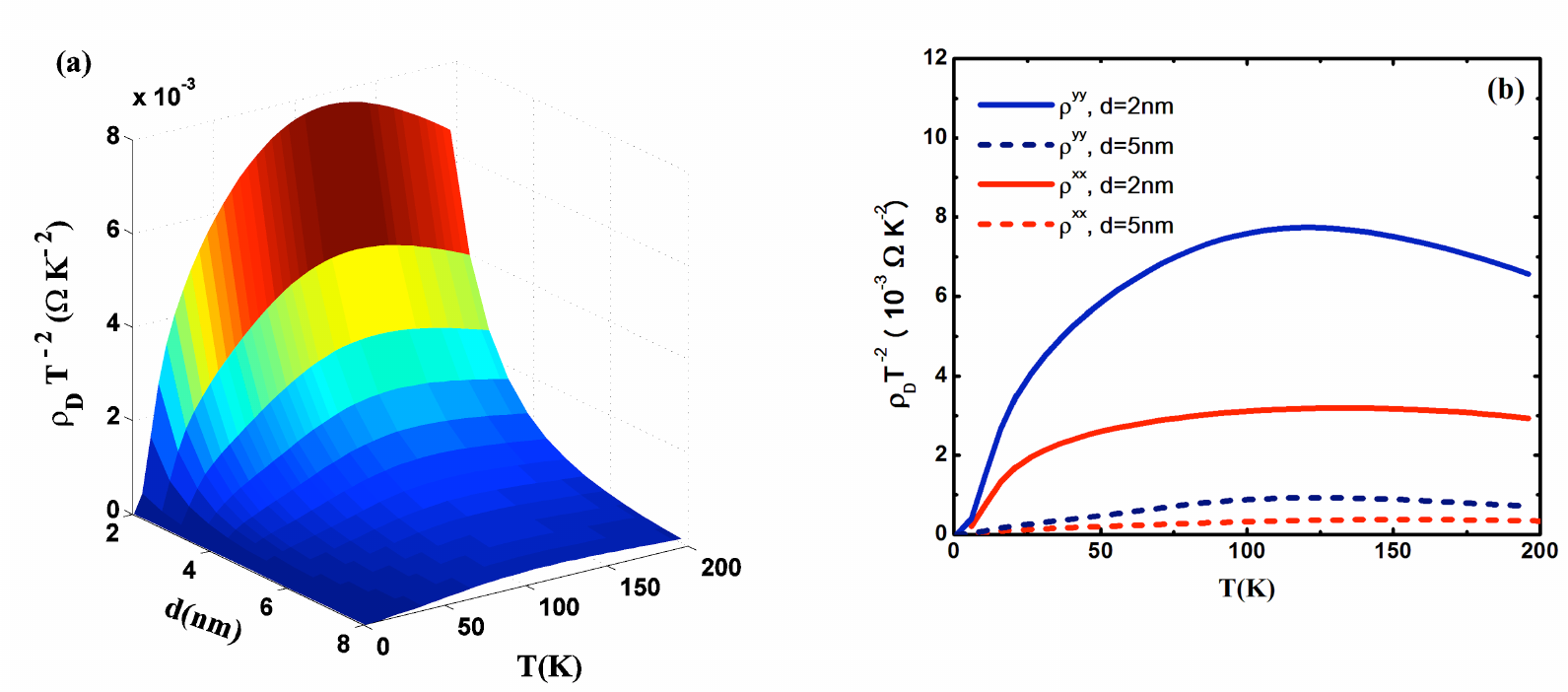}
	\caption{(Color online) Scaled drag resistivity calculated within RPA (a) along $y$ direction as a function of temperature and the distance between two layers and (b) along $x$ and $y$ directions as a function of temperature with two different inter-layer separations. Here, $n = 3\times10^{12}$cm$^{-2}$ and $\eta=$1 meV and system comprising of two aligned phosphorene monolayers sandwiched by Al$_{2}$O$_{3}$.}
	\label{fig7}
\end{figure*}
\begin{figure*}[t]
	\includegraphics[width=16cm]{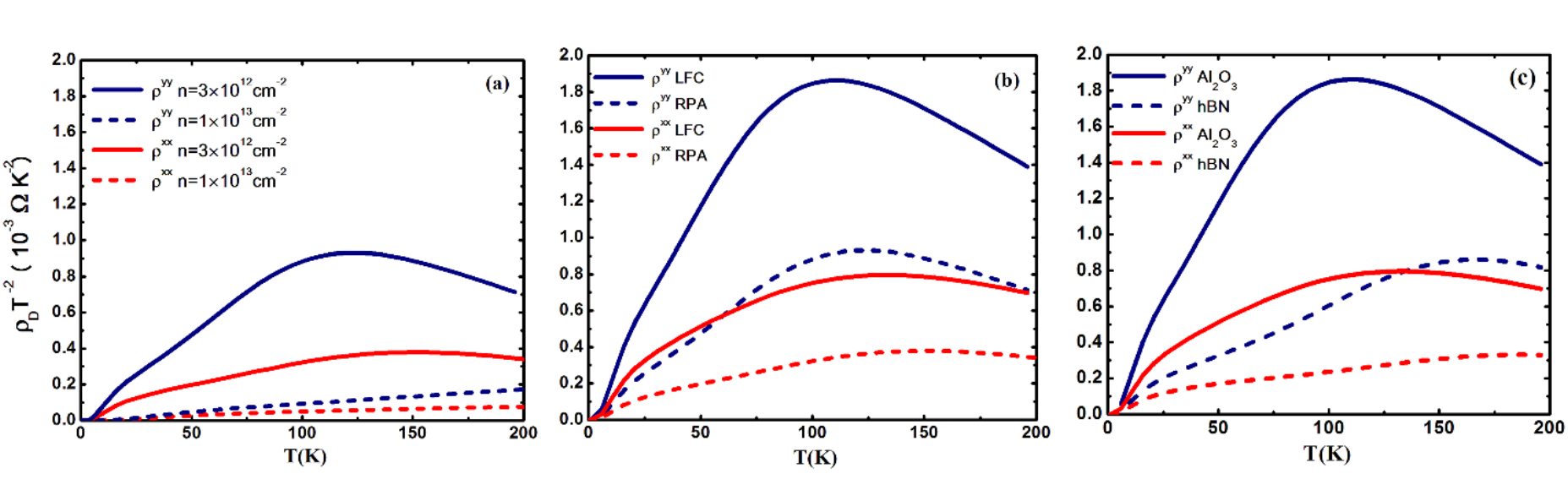}
	\caption{(Color online) Scaled drag resistivity, $\rho_{D}T^{-2}$, as a function of temperature for two aligned phosphorene monolayers sandwiched by Al$_{2}$O$_{3}$ and calculated (a) within RPA at two different electron densities $n = 3\times10^{12}$cm$^{-2}$ (solid line) and $n = 1\times10^{13}$cm$^{-2}$ (dashed line) (b)within RPA and Hubbard local field approximation at electron density $n = 3\times10^{12}$cm$^{-2}$ and (c) sandwiched by Al$_{2}$O$_{3}$ and \BN\ calculated within Hubbard local field approximation at electron density $n = 3\times10^{12}$cm$^{-2}$. Here we set $d=$ 5 nm and $\eta=$ 1 meV.}
	\label{fig8}
\end{figure*}
In order to understand how rotation of one layer with respect to the other about the normal direction to the layers ($z$ direction) impacts the behavior of drag resistivity, we present calculations of the diagonal and off-diagonal elements of the drag resistivity matrix for a couple of rotational angles in Fig. \ref{fig6}. Here, we set $\tau_{1}=0$ and $\tau_{2}=\tau$. It can be seen that as the angle of rotation increases, both diagonal elements of drag resistivity decrease considerably. This observation can be rationalized through the fact that by increasing the angle of rotation, one of the plasmonic branches is forced into the damped regime. As a result, the contribution of plasmons to the Coulomb drag phenomenon, which is known to be significant, decreases. When the angle of rotation is $\pi/2 $, the diagonal components of the drag resistivity tensor have zero values. In this configuration where the \textit{x}-axis of one layer lies along the \textit{y}-axis of the other layer, the values of diagonal elements, $ \rho^{\alpha\alpha} $ , become exactly equal to those of the off-diagonal elements for a system with no rotation. Furthermore, our calculations show that when the layers are rotated with respect to one another the anisotropic effects can create an interesting non-zero transversal drag resistivity, $\rho^{xy}$, which is absent in isotropic materials at zero magnetic field. This observation can be fully understood by Eq.(\ref{eq12}-\ref{eq14}) in which a misalignment of the laboratory and the layer axes gives rise to a non-zero value for the off-diagonal elements. This effect, however, may exist in a double-layer structures subjected to a perpendicularly applied magnetic field \cite{Gramila:prl91,Gorbachev:natphys12}.

Another interesting geometrical effect in a double-layer phosphorene structure comes from changing the inter-layer distance, which is presented in Fig.\ref{fig7}. Fig.\ref{fig7}(a) is a 3D plot showing the variation of $\rho^{yy}T^{-2}$ as a function of layer spacing and temperature. It can be observed that the peaks occur at intermediate temperatures where the plasmon contribution to the drag resistivity is significant, over the whole range of inter-layer distances. Also the scaled drag resistivity decreases strongly when increasing the separation between two layers at all temperatures. One can attribute this behavior to the inter-layer interaction, which decays exponentially with the increasing distance between layers, and decreases due to acoustic modes shifting toward higher energies. Having said that, it is worth pointing out that the changing inter-layer distance does not significantly change the drag anisotropy ratio. To trace this behavior, we plot both scaled $\rho^{xx}$ and $\rho^{yy}$ as a functions of temperature for two different layer spacings: $d=$ 2 nm  and 5 nm in Fig.\ref{fig7}(b), which shows that the anisotropy ratio is less dependent on the inter-layer distance.

The effect of electron density on the drag resistivity is also of interest; hence, we illustrate it in Fig.\ref{fig8}(a). As expected for double-layers systems, for which the electron density increases, the Coulomb drag decreases and the resistivity peak moves toward higher temperature\cite{Flensberg:prl94}. Moreover, it is worth mentioning that the anisotropic effect is more pronounced at lower electron density.
\begin{figure}
	\includegraphics[width=8.5cm]{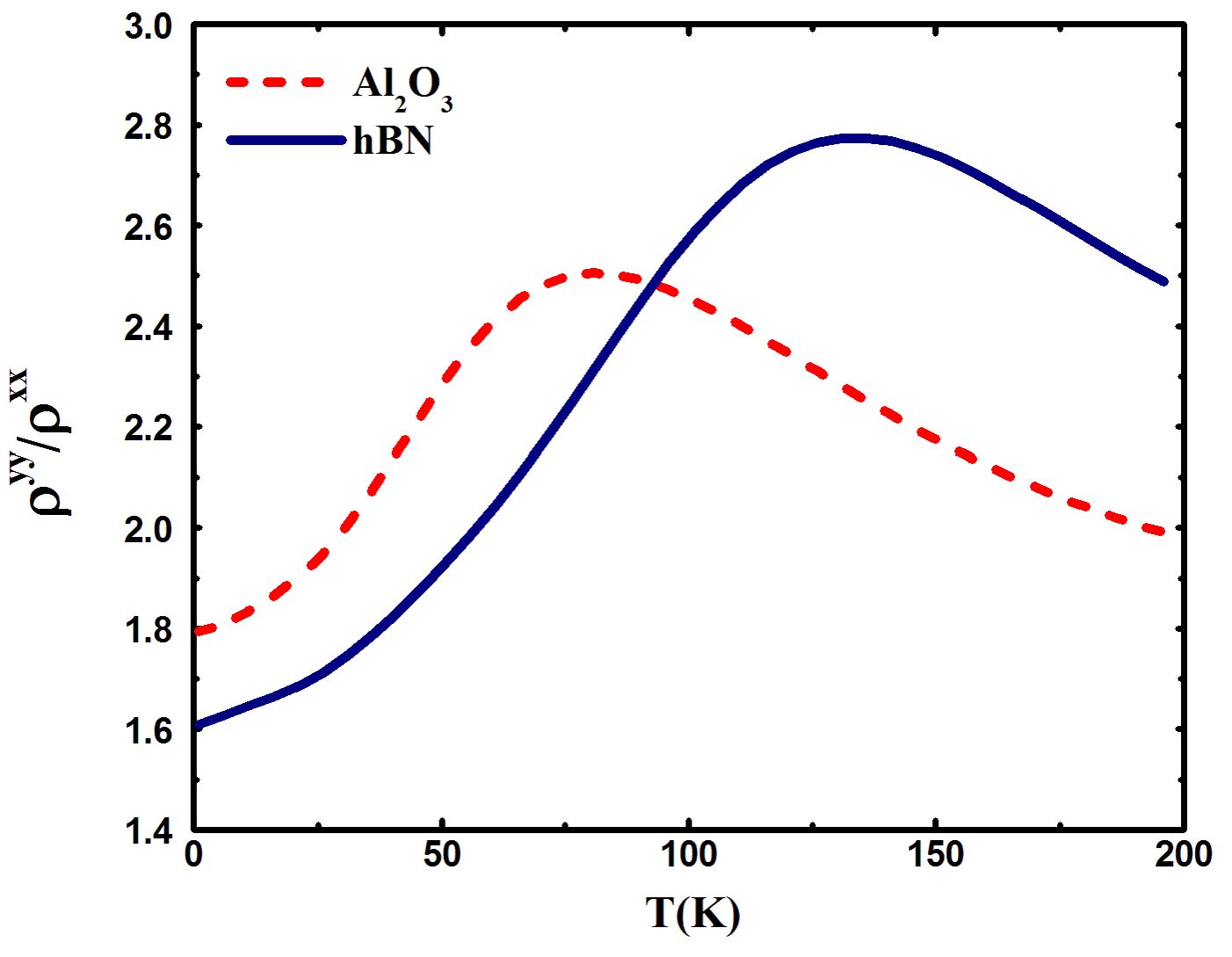}
	\caption{(Color online) The anisotropy ratio of drag resistivity, $\rho^{yy}/\rho^{xx}$, as a function of temperature for two aligned phosphorene monolayers sandwiched by Al$_{2}$O$_{3}$ (solid line) and \BN\ (dashed line) calculated within Hubbard local field approximation at electron density $n = 3\times10^{12}$cm$^{-2}$. Here we set $d=$ 5 nm and $\eta=$ 1 meV.}
	\label{fig9}
\end{figure}
By including the Hubbard zero-temperature LFC, improvements on the RPA results are studied in Fig.\ref{fig8}(b). Here, we employ the intra-layer local field factor, Eq.(\ref{eq5}), which is responsible for most of the drag resistivity enhancement by the short-range interaction effects \cite{swierkowski:prl95}. Exchange interaction, which is taken into account by the Hubbard LFC, impacts the inter-layer interaction through the dielectric function of the system (see Eq.(\ref{eq3})). Our calculations show that including the LFC factor enhances the drag resistivity results notably by strengthening the inter-layer interaction \cite{swierkowski:prl95}. Furthermore, for the parameters used here one can see that the values of the anisotropy ratio are slightly different for both approximations.
 We also investigate the effects of substrate and spacer dielectric materials on anisotropic drag resistivity by considering two already experimentally used insulators, namely Al$_{2}$O$_{3}$ \cite{Luo:ieee14} and $h$-BN \cite{Xiaolong:natc15}, in phosphorene systems. Here, we assume $n = 3\times10^{12}$cm$^{-2}$ corresponding in \BN\ case to density parameter $r_{s}\sim5$ which makes it necessary to consider the LFC factor in our calculation. The results indicate that including the LFC in drag resistivity calculations is important for the both studied substrates. Significant quantitative differences between the results of RPA and local field factor approximation suggest a strong sensitivity of the drag resistivity to the effective intra-layer electron-electron interactions. Our results indicate that the anisotropic drag resistivity has higher values at all temperatures, when Al$_{2}$O$_{3}$ is used as spacer/substrate compared with the case in which \BN\ is used (see Fig.\ref{fig8}(c)), due to a larger dielectric constant of Al$_{2}$O$_{3}$. Employing the high- $\kappa$ materials as substrates/spacers enhances the inter-layer electron-electron interaction due to the reduced screening effects between two layers. 
  
Finally, we present the anisotropy ratio for two different substrates in Fig. \ref{fig9}. We have employed Hubbard LFC to the dielectric function to account for the exchange short-range effects. Calculations show that different substrates have slightly different effects on the anisotropy ratio and shift the maximum expected anisotropy ratio.


\section{Conclusion}
\label{sec:final}
To summarize, we have derived a formula for the  drag resistivity in a structure composed of two spatially separated 2D electron gas systems with anisotropic parabolic band structures. We have assumed the electron gases are sandwiched by insulators so that there is no tunneling between the layers. We have chosen double-layer phosphorene as an example system on which we apply our anisotropic drag theory. Our numerical results confirm that the drag resistivity depends not only on the typically considered parameters such as temperature, inter-layer separation, carrier density and nature of elementary excitations, but also on the direction of momentum transfer between the two layers in addition to the rotational parameter. Our calculations also show that while the diagonal elements of anisotropic drag resistivity tensor have different values due to different electron effective masses along $x$ and $y$ directions at any temperatures of interest, there are non-zero off-diagonal elements for the rotated structure. The non-zero off-diagonal elements have not been reported before in a 2D coupled system without an applied magnetic field. According to the numerical results, both diagonal elements of anisotropic drag resistivity tensor increase with decreasing inter-layer separation and electron density. We show that, the anisotropic ratio varies effectively with the change of temperature and electronic density. To improve on RPA results at low electron density, we have included the zero temperature Hubbard LFC factor in our calculations and shown that the inclusion of LFC enhances the drag resistivity values by almost a factor of 2. Moreover, we have studied the effects of substrate/spacer on the anisotropic drag resistivity. We show that a substrate/spacers with high dielectric constant is able to increase anisotropic drag resistivity considerably.

These results provide qualitative insight into the impact that anisotropic band structure can have on drag resistivity, as an important transport quantity in a coupled 2D structure. The present work also suggests that the rotational parameter between layers can be considered as an extra degree of freedom for the applications of momentum transfer between coupled layers.


\acknowledgments

It is a pleasure to thank D. Otalvaro for making useful comments on this manuscript.

	\appendix
	\section{\bf{Drag Resistivity Tensor In A Double-Layer System With Anisotropic Parabolic Band Structure}}

		Here, we present a derivation of Eq.(\ref{eq:mod2}) for the drag resistivity tensor in a rotationally misaligned double-layer electron gas system with anisotropic parabolic band structure by following the approach of Ref. \cite {Flensberg:prb95a}, closely. We suppose that the intra-layer transport time is independent of the wave vector. So, in a 2D system with  energy dispersion of the form of Eq.(\ref{eq:energy}), the effective mass, $\hat{M} $, transport time, $ \hat{\tau}_{t} $, and mobility, $\hat{\mu}_{t}$, are symmetric and momentum-independent 2$\times$2 tensors and related by:
		\begin{equation}
		\hat{\mu}_{t}=e\hat{M}^{-1}\hat{\tau}_{t}
		\label{A1}
		\end{equation}
		 We assume one layer (layer 1) is fixed and take its lattice principal axes along the laboratory coordinate system but the other layer (layer 2) is rotated by an angle $ \tau $. Hence, in the laboratory frame, $\hat{M} $, $ \hat{\tau}_{t} $, and  $\hat{\mu}_{t}$ have zero and non-zero off-diagonal elements in the fixed and rotated layers, respectively. In layer 2, these non-diagonal matrices can be expressed in terms of the diagonal ones by introducing rotation matrix, $ \hat{R}(\tau) $, \textit{e.g.} for effective mass tensor we have
		
		\begin{equation}
		\hat{M}_{2}=\hat{R}(-\tau)\hat{M}_{1}\hat{R}(\tau)
		\label{A2}
		\end{equation}
		Since we are dealing with the symmetric effective mass and transport time tensors, each one is equal to its transpose. Moreover, due to  the diagonal representation of these tensors in the laboratory frame, we have:
		\begin{equation} \hat{M}_{i}^{-1}\hat{\tau}_{t,i}=\hat{\tau}_{t,i}\hat{M}_{i}^{-1}
		\label{A3} 
		\end{equation} 
		
		In the Boltzmann transport equation framework, we define a deviation function $ g(\mathbf{k}) $ as
		
		\begin{equation}  
		\delta f\equiv f(\mathbf{k})-f^{0}(\mathbf{k}) =-k_{B}T \bigg(\frac{\partial f^{0}(\mathbf{k})}{\partial E_{\mathbf{k}}}\bigg) g(\mathbf{k})
		\label{A4}
		\end{equation}
		where $ f(\mathbf{k}) $ is the non-equilibrium Fermi distribution function and
		$ f^{0}(\mathbf{k})=f^{0}(E_{\mathbf{k}}) $. The linearized inter-layer collision integral is given by:
		
		\begin{equation}
		\begin{aligned}
		\begin{split}
		S[g_{1},g_{2}](\mathbf{k_{2}}) &=2 \int\frac{d\mathbf{k_{1}}}{(2\pi)^{2}}\int\frac{d\mathbf{q}}{(2\pi)^{2}} w(\mathbf{q}, E_{\mathbf{k_{1}+q}}-E_{\mathbf{k_{1}}})\\
		&\times
		f^{0}_{1}(\mathbf{k_{1}})f^{0}_{2}(\mathbf{k_{2}})
		\big[1-f^{0}_{1}(\mathbf{k_{1}+q})\big]\\
		&\times\big[1-f^{0}_{2}(\mathbf{k_{2}-q})\big] \big[g_{1}(\mathbf{k_{1}})+g_{2}(\mathbf{k_{2}})\\
		&- g_{1}(\mathbf{k_{1}+q})-g_{2}(\mathbf{k_{2}-q})\big]\\
		&\times
		\delta(E_{\mathbf{k_{1}}}+E_{\mathbf{k_{2}}}-E_{\mathbf{k_{1}+\mathbf{q}}}-E_{\mathbf{k_{2}-\mathbf{q}}}) 
		\label{A5}
		\end{split}
		\end{aligned}
		\end{equation}
		with
		
		\begin{equation}  
		w(\mathbf{q},\omega)=4\pi\hbar^{-1}|U_{21}(\mathbf{q},\omega)|^{2}
		\label{A6}  
		\end{equation}
		
		Considering weak inter-layer interaction, the coupled Boltzmann equations are given by: 
		
		\begin{equation}  
		e_{1}(\frac{\partial f^{0}_{1}}{\partial E})(\mathbf{v_{1}})^{t}.\mathbf{\Xi_{1}}=-\hat{H}_{1}[g_{1}](\mathbf{k_{1}})
		\label{A7}
		\end{equation}
		and
		\begin{equation} 
		e_{2}(\frac{\partial f^{0}_{2}}{\partial E})(\mathbf{v_{2}})^{t}.\mathbf{\Xi_{2}}=-\hat{H}_{2}[g_{2}](\mathbf{k_{2}})+S[g_{1},g_{2}=0](\mathbf{k_{2}})
		\label{A8}
		\end{equation}     
		where the superscript $ t $ means the transpose and $\mathbf{\Xi_{i}}$ and $ \hat{H}_{i} $ are the electric field and negative of the linearized intra-layer collision operator in layer \textit{i}, respectively. In a 2D semiconductor with anisotropic parabolic band structure, the electron velocity is simply related to the wave vector 
		\begin{equation} \mathbf{v_{i}}(\mathbf{k_{i}})=\hbar\hat{M}_{i}^{-1}\mathbf{k_{i}}
		\label{A9}
		\end{equation}
		From the above equations 
		$ g_{1}$ and $g_{2} $ can be obtained as:
		\begin{equation}  
		g_{1}(\mathbf{k_{1}})=-e_{1}\hat{H}_{1}^{-1}\big[(\frac{\partial f^{0}_{1}}{\partial E})(\mathbf{v_{1}})^{t}\big](\mathbf{k_{1}}).\mathbf{\Xi_{1}}
		\label{A10}
		\end{equation}
		and
		\begin{equation} 
		g_{2}(\mathbf{k_{2}})=-e_{2}\hat{H}_{2}^{-1}\big[(\frac{\partial f^{0}_{2}}{\partial E})(\mathbf{v_{2}})^{t}\big](\mathbf{k_{2}}).\mathbf{\Xi_{2}}+\hat{H}_{2}^{-1}[S](\mathbf{k_{2}})
		\label{A11}
		\end{equation} 
		Since the current in layer 2 is equal to zero
		
		\begin{equation} 
		\mathbf{J_{2}}=-2e_{2}k_{B}T\int\frac{d\mathbf{k_{2}}}{(2\pi)^{2}}\bigg((\frac{\partial f^{0}_{2}}{\partial E})\mathbf{v_{2}}\bigg)g_{2}(\mathbf{k_{2}})=0
		\label{A12}
		\end{equation} 
		one finds the following relation:
		\begin{equation} 
		\begin{split}
		&2e_{2}^{2}k_{B}T\int\frac{d\mathbf{k_{2}}}{(2\pi)^{2}}\bigg((\frac{\partial f^{0}_{2}}{\partial E})\mathbf{v_{2}}\bigg)\hat{H}_{2}^{-1}\big[(\frac{\partial f^{0}_{2}}{\partial E})(\mathbf{v_{2}})^{t}\big](\mathbf{k_{2}}).\mathbf{\Xi_{2}}\\		
		&=-2e_{2}k_{B}T\int\frac{d\mathbf{k_{2}}}{(2\pi)^{2}}\bigg((\frac{\partial f^{0}_{2}}{\partial E})\mathbf{v_{2}}\bigg)\hat{H}_{2}^{-1}[S](\mathbf{k_{2}})
		\label{A13}
		\end{split}
		\end{equation}
		The left hand side of the above equation is equals to $n_{2}e_{2}\hat{\mu}_{t,2}.\mathbf{\Xi_{2}}  $. By employing the following identities to Eq.(\ref{A5})
	
		\begin{equation} 
		\begin{split}
		&\delta(E_{\mathbf{k_{1}}}+E_{\mathbf{k_{2}}}-E_{\mathbf{k_{1}+\mathbf{q}}}-E_{\mathbf{k_{2}-\mathbf{q}}})\\&=\hbar\int_{0}^{\infty}d\omega \delta(E_{\mathbf{k_{1}}}-E_{\mathbf{k_{1}+q}}-\hbar\omega)\delta(E_{\mathbf{k_{2}}}-E_{\mathbf{k_{2}-q}}+\hbar\omega)\\
		\label{A14}
		\end{split}
		\end{equation}
		and 
		\begin{equation} 
		f^{0}(E_{1})[1-f^{0}(E_{2})]=[f^{0}(E_{2})-f^{0}(E_{1})]n_{B}(E_{1}-E_{2})
		\label{A15}
		\end{equation} 
		
		Eq.(\ref{A13}) can be rewritten as:  
		
		\begin{equation}
		\begin{aligned}
		\begin{split}
		n_{2} e_{2} \hat{\mu}_{t,2}.\mathbf{\Xi}_{2}&=-\frac{4e_{2}k_{B}T}{\pi}\int\frac{d\mathbf{q}}{(2\pi)^{2}} \int_{0}^{\infty}d\omega |U_{21}(\mathbf{q},\omega)|^{2}\\
		&\times n_{B}(\hbar\omega) n_{B}(-\hbar\omega)\\ 	
		&\times \bigg[ \int\frac{d\mathbf{k_{2}}}{(2\pi)^{2}} [{f^{0}_{2}(\mathbf{k_{2}})-f^{0}_{2}(\mathbf{k_{2}}+\mathbf{q})}]\\
		&\times \delta(E_{\mathbf{k_{2}}}-E_{\mathbf{k_{2}+\mathbf{q}}}-\hbar \omega)\hat{H}^{-1}_{2}\big[(\frac{\partial f^{0}_{2}}{\partial E})\mathbf{v_{2}}\big](\mathbf{k_{2}}) \bigg] 
		\\
		&\times
		\bigg[ \int\frac{d\mathbf{k_{1}}}{(2\pi)^{2}}        [f^{0}_{1}(\mathbf{k_{1}})-f^{0}_{1}(\mathbf{k_{1}}+\mathbf{q})]\\
		&\times \delta(E_{\mathbf{k_{1}}}-E_{\mathbf{k_{1}+\mathbf{q}}}-\hbar \omega)[g_{1}(\mathbf{k_{1}})-g_{1}(\mathbf{k_{1}}+\mathbf{q})] \bigg]          	   			    			\label{A16}
		\end{split}
		\end{aligned}
		\end{equation}
		
		where $n_{B}(\hbar\omega)$ is the  Bose distribution function and in the second line we use the Hermitian property of $ H_{2}^{-1} $. Using relaxation time approximation, one can define: 
		\begin{equation}  
		g_{i}(\mathbf{k_{i}})= -e_{i}\hat{H}_{i}^{-1}\big[(\frac{\partial f^{0}_{i}}{\partial E})(\mathbf{v_{i}})^{t}\big](k_{i}).\mathbf{\Xi_{i}}\equiv  \frac{e_{i}{\big(\hat{\tau}_{t,i}\mathbf{v_{i}}(\mathbf{k_{i}})\big)}^{t}.\mathbf{\Xi_{i}}}{k_{B}T}
		\label{A17}
		\end{equation}
		Hene, Eq.(\ref{A16}) can be rewritten as:
		
		\begin{equation}
		\begin{split}
		n_{2} e_{2} \hat{\mu}_{t,2}.\mathbf{\Xi}_{2}&=-\frac{4e_{2}k_{B}T}{\pi}\int\frac{d\mathbf{q}}{(2\pi)^{2}} \int_{0}^{\infty}d\omega |U_{21}(\mathbf{q},\omega)|^{2}\\
		&\times n_{B}(\hbar\omega) n_{B}(-\hbar\omega)\\ 	
		&\times         	
		\bigg[\frac{1}{2k_{B}T}\int\frac{d\mathbf{k_{2}}}{(2\pi)^{2}}[{f^{0}_{2}(\mathbf{k_{2}})-f^{0}_{2}(\mathbf{k_{2}}+\mathbf{q})}]\\
		&\times \delta(E_{\mathbf{k_{2}}}-E_{\mathbf{k_{2}+\mathbf{q}}}-\hbar \omega)\hat{\tau}_{t,2}[\mathbf{v}_{2}(\mathbf{k_{2}}+\mathbf{q})\\
		&-\mathbf{v}_{2}(\mathbf{k_{2}}	)] \bigg]
		\bigg[\frac{e_{1}}{k_{B}T} \int\frac{d\mathbf{k_{1}}}{(2\pi)^{2}}[f^{0}_{1}(\mathbf{k_{1}})\\
		&-f^{0}_{1}(\mathbf{k_{1}}+\mathbf{q})] \delta(E_{\mathbf{k_{1}}}-E_{\mathbf{k_{1}+\mathbf{q}}}-\hbar \omega)\\
		&\times \bigg(\hat{\tau}_{t,1} 
		[\mathbf{v}_{1}(\mathbf{k_{1}}+\mathbf{q})-\mathbf{v}_{1}(\mathbf{k_{1}}	)]\bigg)^{t} .\mathbf{\Xi}_{1} \bigg]
		\label{A18}
		\end{split}
		\end{equation}

		The DC electric fields in two layers are related by: 	 
		\begin{equation}
		\mathbf{\Xi}_{2}=\hat{\rho}_{21}\mathbf{J}_{1}
		=n_{1} e_{1}\hat{\rho}_{21} \hat{\mu}_{t,1}.\mathbf{\Xi}_{1}
		\label{A19}
		\end{equation}
		Inserting Eqs.(\ref{A9}) and (\ref{A19}) into Eq.(\ref{A18}) and considering the properties of effective mass and transport time tensors, one gets:
		
       \begin{widetext}
		\begin{equation}
		\begin{split}
		n_{1} e_{1}n_{2} e_{2} \hat{\mu}_{t,2}\hat{\rho}_{21} \hat{\mu}_{t,1}.\mathbf{\Xi}_{1}&= -\frac{4e_{2}k_{B}T}{\pi}\int\frac{d\mathbf{q}}{(2\pi)^{2}} \int_{0}^{\infty}d\omega |U_{21}(\mathbf{q},\omega)|^{2} n_{B}(\hbar\omega) n_{B}(-\hbar\omega)\\ 	
		&\times         	
		\bigg[\frac{\hat{\tau}_{t,2}\hat{M}_{2}^{-1}\hat{R}(-\tau)\mathbf{q} }{2k_{B}T}\int\frac{d\mathbf{k_{2}}}{(2\pi)^{2}}[{f^{0}_{2}(\mathbf{k_{2}})-f^{0}_{2}(\mathbf{k_{2}}+\mathbf{q})}] \delta(E_{\mathbf{k_{2}}}-E_{\mathbf{k_{2}+\mathbf{q}}}-\hbar \omega)   \bigg]\\
		&\times
		\bigg[\frac{\mathbf{q}^{t}e_{1}\hat{M}_{1}^{-1}\hat{\tau}_{t,1}.\mathbf{\Xi}_{1}}{k_{B}T} \int\frac{d\mathbf{k_{1}}}{(2\pi)^{2}}[f^{0}_{1}(\mathbf{k_{1}})-f^{0}_{1}(\mathbf{k_{1}}+\mathbf{q})] \delta(E_{\mathbf{k_{1}}}-E_{\mathbf{k_{1}+\mathbf{q}}}-\hbar \omega) 
		\bigg]
		\label{A20}
		\end{split}
		\end{equation}
		\end{widetext}
		By multiplying both sides of above equation by $\hat{\mu}^{-1}_{t,2}$ and using mobility relation, Eq.(\ref{A1}), commutative property, Eq.(\ref{A17}) and  the equality $4n_{B}(\hbar\omega) n_{B}(-\hbar\omega)=-\sinh^{-2}(\hbar\omega/2k_{B}T)$ , Eq.(\ref{A20}) can be written as:
		\begin{equation}
		\begin{split}
		{\rho}_{21}^{\alpha \beta}= \frac{\hbar^{2}}{2\pi n_{1} e_{1}n_{2} e_{2} k_{B}T} \int\frac{d\mathbf{q}}{(2\pi)^{2}}q_{\alpha}q_{\beta} \int_{0}^{\infty}d\omega \frac{|U_{21}(\mathbf{q},\omega)|^{2}} {\sinh^{2}(\hbar\omega/2k_{B}T)} \\
		\times \bigg[ \int\frac{d\mathbf{k_{1}}}{(2\pi)^{2}}     [f^{0}_{1}(\mathbf{k_{1}})-f^{0}_{1}(\mathbf{k_{1}}+\mathbf{q})] \delta(E_{\mathbf{k_{1}}}-E_{\mathbf{k_{1}+\mathbf{q}}}-\hbar \omega) \bigg]\\         	
		\times \bigg[\int\frac{d\mathbf{k_{2}}}{(2\pi)^{2}}   [f^{0}_{2}(\mathbf{k_{2}})-f^{0}_{2}(\mathbf{k_{2}}+\mathbf{q})] \delta(E_{\mathbf{k_{2}}}-E_{\mathbf{k_{2}+\mathbf{q}}}-\hbar \omega) \bigg]	
		\label{A21}
		\end{split}
		\end{equation}
	
		where $ q_{\alpha} $ and $ q_{\beta} $ are the $ \alpha $ and $ \beta $ components of transferred wave vector corresponding to the layer 1 and layer 2 in the laboratory frame, respectively. Finally by using the polarizabilion function definition, Eq.(\ref{eq6}), one obtains the Eq.(\ref{eq:mod2}).

%

\end{document}